\documentclass[12pt]{article}

\usepackage[draft]{hyperref}
\usepackage{amsmath}
\usepackage{graphicx}
\usepackage{authblk}

\begin{document}

\title{Intra-cavity gain shaping of mode-locked Ti:Sapphire laser oscillations}

\author{Shai Yefet, Na'aman Amer, and Avi Pe'er$^{*}$}

\affil{Department of physics and BINA Center of nano-technology, Bar-Ilan university, Ramat-Gan 52900, Israel

$^*$e-mail: avi.peer@biu.ac.il}

\date{Dated: \today}

\maketitle

\begin{abstract}
The gain properties of an oscillator strongly affect its behavior. When the gain is homogeneous, different modes compete for gain resources in a `winner takes all' manner, whereas with inhomogeneous gain, modes can coexist if they utilize different gain resources. We demonstrate precise control over the mode competition in a mode locked Ti:sapphire oscillator by manipulation and spectral shaping of the gain properties, thus steering the competition towards a desired, otherwise inaccessible, oscillation. Specifically, by adding a small amount of spectrally shaped inhomogeneous gain to the standard homogeneous gain oscillator, we selectively enhance a desired two-color oscillation, which is inherently unstable to mode competition and could not exist in a purely homogeneous gain oscillator. By tuning the parameters of the additional inhomogeneous gain we flexibly control the center wavelengths, relative intensities and widths of the two colors.
\end{abstract}

\section{Introduction}

Active oscillators, and in particular laser oscillators, which produce precise, stable oscillations are a major concept in physics and engineering.
Due to the highly nonlinear interplay between positive feedback, loss, gain saturation (and possibly additional nonlinear effects), the steady state solution of a laser oscillation involves strong competition over gain resources between all possible modes of oscillation \cite{Siegman@Lasers}. The ability to control mode competition and manipulate the oscillating cavity modes, stands at the heart of laser physics and applications.

The properties of the gain play a central role in the dynamics of a laser oscillator and in particular in determining the steady state oscillation. In textbooks of laser physics, gain is classified as either homogeneously or inhomogeneously broadened \cite{Yariv@Quantum_electronics}. The effect of mode competition takes place in homogeneous broadening, where one global gain resource (population inversion in all the gain atoms) is accessible to all oscillation frequencies within the gain bandwidth of the active medium, causing the mode with the highest net gain to dominate over all the other possible modes. Consequently, in CW operation, a laser with homogeneous gain would inherently tend towards single mode operation. With inhomogeneous broadening however, each mode has its own gain resource and different frequencies in the oscillator do not compete. Consequently, a laser with inhomogeneously broadened gain would tend to multimode operation, with an oscillation spectrum that reflects the net gain spectrum. By shaping the loss profile one can enforce narrow oscillations on an inhomogeneously broadened laser at a cost of reduced pumping efficiency according to the ratio of the actual oscillation bandwidth to the bare gain bandwidth. For example, dual frequency oscillation can be obtained with inhomogeneously broadened gain by proper loss shaping and filtering only the desired frequencies (with an efficiency cost). With homogeneous gain however, only the identity of the single winning mode can be affected by loss shaping, but dual frequency oscillation cannot be enforced.

For a mode locked oscillation, an additional factor comes into play. First, the nonlinear loss imposed by the mode locking mechanism forces broadband, phase synchronized oscillations. Second, the spectrum is dictated by the delicate balance between gain, loss, dispersion profile and the temporal response of the mode locking mechanism. Due to these factors,the typical result is a pulse with a broad, smooth, single band spectrum. The situation with regard to intra-cavity spectral shaping however, is similar to CW. With inhomogeneous gain, multi-color operation is possible, as was demonstrated with mode locked semiconductor lasers near threshold \cite{Delfyett@Semiconductor_modelocked}, whereas for homogeneous gain, loss shaping can only set the allowed bands of oscillation, but between these bands, mode competition will usually choose one final winner oscillating band. Consequently, dual color modelocked oscillation is inherently unstable in a homogeneously broadened laser, and can be achieved only if the two colors have similar gain.

It so happens that most common mode locked lasers are primarily homogeneously broadened, mainly because the efficiency of pump utilization is higher for homogeneous gain. Even semiconductor lasers, which are inhomogeneously broadened near threshold, tend to become homogeneous as pumping is increased \cite{Homogenoues_semiconductor1,Homogenoues_semiconductor2}. Thus, if high power, dual color (or more) oscillation is desired, one must find a way to overcome mode competition in the homogeneous gain. We note that obtaining multi-color oscillation is more than an interesting exercise in laser physics; A multi-color (in particular dual color) oscillation is necessary for important applications, such as Raman spectroscopy \cite{antistokes}, Raman microscopy \cite{twophoton,sampleimaging}, and direct frequency comb spectroscopy \cite{combspec,ramandynamics}.

Many attempts were performed in the past to obtain dual color mode locking. For example, dual lobed loss filtering \cite{doubleslit,threeprisms} or dual output coupling \cite{twoOCs} was attempted with minimal success, as these are inherently loss shaping techniques, that do not address the problem of mode competition. Other attempts bypassed the mode competition problem using active or passive synchronization of two independent lasers \cite{phaselocking,synchronization}, either by coupling two separate cavities through a shared gain medium \cite{twobeams,fourprisms}, or by synchronously pumping two OPOs \cite{OPOs}. All of these methods require several oscillators and special care for stabilization of timing jitter between the participating pulse trains.

\begin{figure}
\centering\includegraphics[width=10cm]{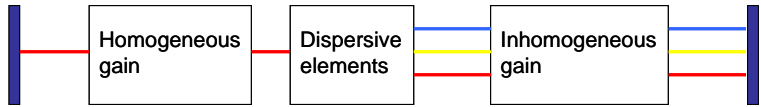}
\caption{\label{diagram} Block diagram illustrating the core principle of intra-cavity gain shaping.}
\end{figure}

Here we demonstrate a method to directly control the mode competition in a \emph{single} oscillator, steering it towards the desired dual-color oscillation. The core principle is to tailor the gain profile instead of the loss inside the optical cavity. In addition to the homogeneous gain medium, we place a 2nd gain medium at a position in the cavity where the spectrum is spatially dispersed (as schematically shown in Fig. \ref{diagram}). In this position, different frequencies pass at different physical locations in the gain medium and therefore do not compete for gain. Furthermore, by proper spatial shaping of the pump beam in this additional gain medium one can shape the spectral gain profile. As opposed to the first homogeneously broadened gain medium, this additional gain is inherently inhomogeneous with a spectral shape of our desire. In this method, any combination of homogeneous and inhomogeneous gain can be realized by varying the splitting ratio of pump power between the two gain media. While most of the gain remains in the standard homogeneous medium, the additional inhomogeneous gain allows us to enhance specific frequencies in the overall spectrum by selectively adding gain to these frequencies, thus boosting them in the overall mode competition for the homogeneous gain. As demonstrated here on, this method allows steering the oscillation towards the desired double lobed (or more) spectrum, and shaping of the pulse spectrum almost at will, while preserving the total pulse power, and with minimal added pump power. We note that the concept of passively mode-locked laser with a single gain medium in the dispersive arm was introduced before \cite{inhomogeneousgain}. However, using only inhomogeneous gain is highly inefficient in pump energy, as it requires pumping of a much larger volume. It is therefore much more efficient add only small amount of inhomogeneous gain in order to steer the competition in the homogeneous gain towards the desired oscillations.

\section{Experimental and results}

\begin{figure}
\begin{center}
\includegraphics[width=10cm]{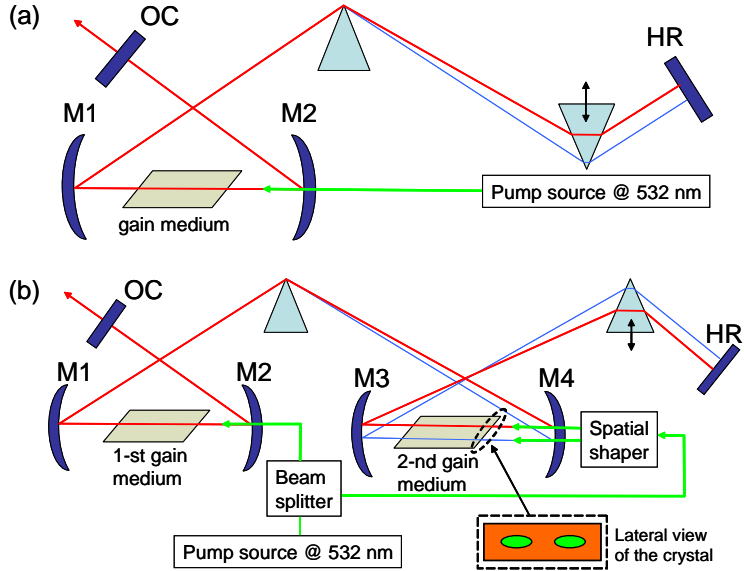}
\caption{\label{standardandnoveldesign} (a) Schematic of the standard design of a
TiS oscillator. A linear cavity composed of a TiS crystal (Ti:Al$_{2}$O$_{3}$) as gain medium placed
between two focusing curved mirrors (M1, M2), and a prism pair (or chirped mirrors) for dispersion compensation. (b) Schematic of the intra-cavity shaped oscillator. A 2nd TiS gain medium is placed at the Fourier plane of a $%
1\times 1$ telescope placed between the prisms (Both TiS crystals were $3$ $mm$ long, $0.25$ wt$\%$ doped). The telescope is comprised
of two curved mirrors (M3, M4) of equal focal length $f=100$ $mm$. Since the spectrum
is spatially dispersed in the 2nd gain medium (each frequency component
traverses at a different position), mode competition is canceled resulting
in the ability to tailor the gain profile inside the oscillator by
controlling the spatial shape of the pump in the 2nd gain medium. The inset shows a lateral view of the two pump spots in the 2nd gain medium.}
\end{center}
\end{figure}

The standard design of a linear Ti:Sapphire (TiS) oscillator with a homogeneous gain medium is illustrated in Fig. \ref{standardandnoveldesign}(a), where a prism pair is commonly used to control dispersion \cite{prismpair}. Due to the homogenous gain and without any loss shaping, the CW operation of the standard design is very narrow band, and the modelocked operation is characterized by a pulse with a single band, broad, smooth spectrum. In our novel design, presented in Fig. \ref{standardandnoveldesign}(b), a unity magnification telescope is inserted between the prisms, with a second TiS crystal as an additional gain medium placed at the Fourier plane of the telescope. This effectively
forms an intra-cavity pulse shaper, where it is possible to pump individual frequency components, and control the spectral amplitude of the oscillating modes.

\begin{figure}
\begin{center}
\includegraphics[width=8.5cm]{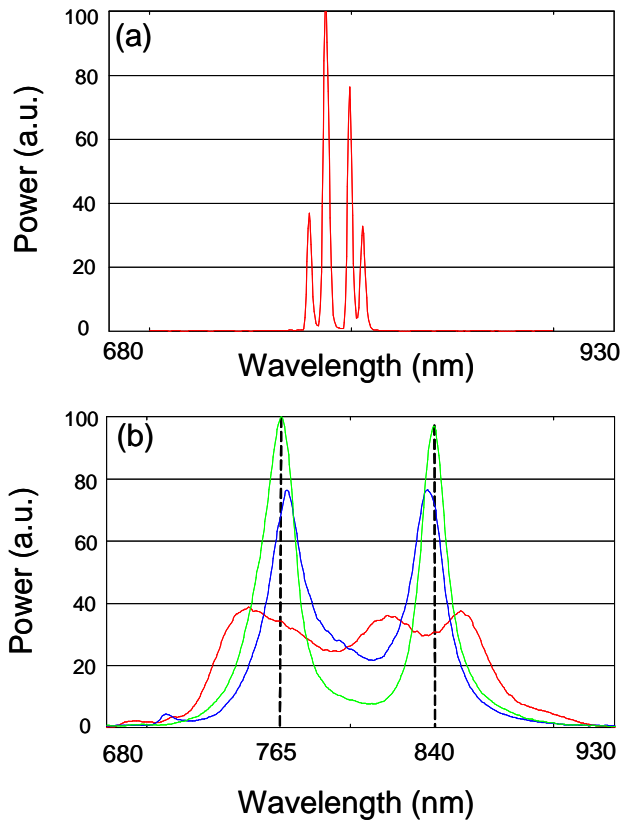}
\caption{\label{CWpulse} Spectra of CW and pulsed operation of the cavity. (a) CW spectrum demonstrating cancelation of mode
competition by the coexistence of multiple CW modes (fingers) when pumping
only the 2nd medium with an elliptically shaped pump spot. In this study, the prisms was made of BK7 glass with dispersive power of $d\theta /d\lambda =0.04$ $rad/\mu m$, @ $\lambda =0.8$ $\mu m$. Given a mode diameter of $21$ $\mu m$ the resolution of the intra-cavity shaper is $9.3$ nm (the bandwidth occupied by a single mode on the surface of the 2nd gain medium). We used cylindrical optics to obtain an elliptically shaped pump beam of $21$ $\mu m$ $\times$ $85$ $\mu m$ at the 2nd medium and we observed $4$ CW fingers that span a bandwidth of $\approx 35$ $nm$, in good agreement with the expected $37$ nm based on the above resolution. (b) Pulsed
spectra observed in the cavity at different stages of pump transfer
from the 1st medium (homogeneous gain) to the 2nd medium (spectrally selective gain). The 2nd medium is pumped at two selected frequencies
with a tightly focused pump, resulting in a spectrum with two sharp lobes (red - initial, blue - intermediate, green - final spectrum).}
\end{center}
\end{figure}

The cancelation of mode competition in the 2nd medium is demonstrated in Fig. \ref{CWpulse}(a) by the continuous-wave (CW) operation of the novel cavity. The oscillator was pumped using a frequency-doubled Nd:YVO$_{4}$ laser at $532$ $nm$ (Verdi by \textit{Coherent}) and when only the 2nd medium was pumped with an elliptically shaped pump spot we achieve a ``multiple fingers" CW operation, indicating that different frequency components coexist. By controlling the shape of the pump spot for the additional gain medium, selected frequencies can be pumped simultaneously without mode competition. For a given prism material, these ``fingers" span a bandwidth corresponding to the spatial width of the elliptically shaped pump spot, and can be centered anywhere within the TiS emission spectrum by scanning the pump beam laterally across the 2nd crystal. The number of modes (``fingers") is determined by the ratio of the pump spot size to the resolution of the intra-cavity shaper.

The measured mode-locked operation is depicted in Fig. \ref{CWpulse}(b). First, the laser was mode-locked with only the 1st crystal pumped. Modelocked operation was achieved by standard techniques of mode-locking based on soft aperturing \cite{softaperture} and noise insertion (knocking on one of the prisms) \cite{noiseinsertion}. Using pump power of $3.7W$ and an output coupler of $92\%$, we obtained an average modelocked power of $205mW$ with a broad homogeneous spectrum ($\approx 140$ $nm$ @ FWHM, red curve) and a repetition rate of $\approx 100$ MHz. Pump power was then transferred from the 1st medium to the 2nd medium in several steps. In each step, we add small amounts of power to the overall pumping, and then direct the excess pump power into the 2nd medium, where two specified spots were pumped at positions corresponding to two lobes. During the transfer, gain is increasing only for certain frequency components while decreasing for all other frequencies (blue curve). The final shape of the spectrum has two clear and significant lobes (green curve), one at $765$ $nm$ ($20 $ $nm$ @ FWHM) and the other at $840$ $nm$ ($16$ $nm$ @ FWHM), and the intermediate spectral power drops to $<10\%$ from maximum. During the process of pump transfer the average pulse power remained approximately the same ($205mW$) and the dispersion profile was tuned by translating the second prism in order to compensate dispersion for the desired two colors. The spatial mode of the laser was stable and did not show any significant changes during the entire process of pump transfer. Note that the control of the spectrum inherently requires an increase in pump power, since one must pump (and cross threshold) in a larger volume of the 2nd medium, hence the overall pump power was increased during the process to maintain pulsed operation up to a final level of $5.55W$ which was split between the two media, such that the pump power to the 1st medium dropped to $3.4W$, and $2.15W$ of pump power power was directed into the 2nd medium, further split between the lobes as follows: $1.3W$ to the lobe at $840$ $nm$ and $0.85W$ to the lobe centered at $765$ $nm$. The splitting ratio is affected by the natural
gain as well as dispersion compensation at these wavelengths.

Since the transfer of the pump and hence the shaping of the pulse spectrum was a gradual, adiabatic-like procedure, modelocking was preserved during the entire transfer from the initial broad single band spectrum to the final two-color shape. This indicates that the dual-color spectrum is inherently synchronized in time as a single pulse train, just like the original single band spectrum was. There seem to exist however, an inherent limit on the amount of pump power that can be transferred into the 2nd gain medium. After transferring about half the power (and obtaining a well established two-color spectrum) any attempt to further transfer power causes first the appearance of CW spikes and eventually brakes modelocking. In addition, even when approaching this limit, the intermediate spectrum between the two forming lobes never (and apparently cannot) drops to zero. The reasons for this limit are not fully understood, but it seems as residual homogenous gain is necessary in order to maintain a broadband 'back bone' to connect the two lobes, and to assure that the two colors are synchronized not only in time but also in phase, forming one joint frequency comb, just like the initial unshaped pulse.

\begin{figure}
\begin{center}
\includegraphics[width=8.5cm]{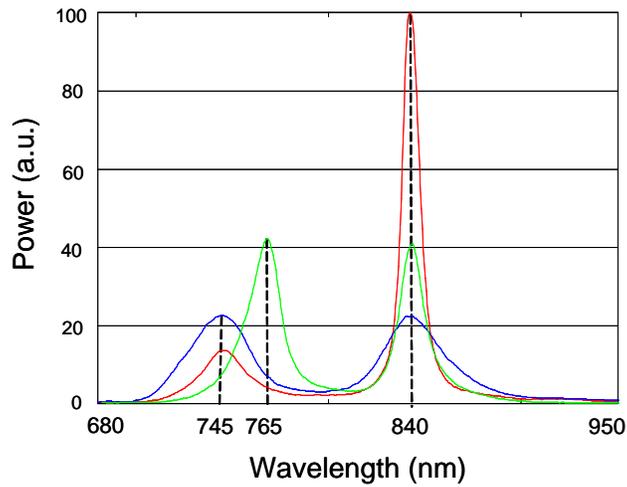}
\caption{\label{cyllinder} Control of spectral power, width and
center position of spectral lobes. Taking a two lobed spectrum as a
reference (green curve): control is demonstrated over the width of each lobe
by changing the spatial width of the pump (blue curve) and the spectral
power of each lobe by changing the power splitting ratio between lobes (red
curve). The center of each lobe is also controlled by sweeping the pump spot
position (left lobe shifted by $20$ $nm$ for both curves to $745$ $nm$).}
\end{center}
\end{figure}

Figure \ref{cyllinder} demonstrates the flexibility to control the spectral
power, width and center wavelength of each lobe. The green curve is that of
Fig. \ref{CWpulse}(b), used here as a reference spectrum. By spatially widening
the pump spots in the 2nd medium the spectral width of each lobe was
increased (blue curve). The spectral power of each lobe was controlled by
adjusting the splitting ratio of the pump power between the spots in the
2nd medium (red curve). Shifting the center position of the lobes by shifting the pump spot laterally is also
demonstrated. The average power is conserved for all curves at $205mW$, and the
intermediate spectral power can be reduced down to several percent only ($<3\%$ from maximum).

A very important feature of our design is that once the spectrum profile
is shaped as desired, mode-locking directly into the
shaped pulse is robust, without the need to repeat the step-by-step pump transfer procedure. We could repeatedly establish stable mode-locking directly into a narrow two-lobed spectrum with $90$ $nm$ separation between
the lobes and intermediate spectral power of $4\%$ from maximum.

In our experiments, the places where lobes can be formed are limited by the spectrum of the initial pulse with pure homogeneous gain before the pump transfer (Fig. \ref{CWpulse}(b), red curve). Trying to pump frequencies beyond the FWHM bandwidth of the initial spectrum resulted in the formation of CW spikes. A broader initial pulse can increase the bandwidth available for gain shaping ,ideally up to the entire emission spectrum of the TiS crystal. The minimum bandwidth of each lobe is limited by the spectral resolution of the intra-cavity shaper. Thus, using prism of higher dispersion, will reduce the bandwidth of the lobes for a given width of the pump beam, but at the same time will increase the pumping volume (and hence the pump power) needed for a given bandwidth of the lobes. The maximum width of the lobes depends only on the pump spatial profile and is limited by available pump power.

So far, the temporal shape of the two-lobed pulse has not been measured. We expect that the envelope of the pulse will have two characteristic time-scales: a short time-scale beating between the two lobes, contained within a long time-scale envelope of the entire pulse. The realization of a measurement system for such structured pulses with broad enough bandwidth on one hand, but with high enough spectral resolution on the other hand is not a simple task and is a future objective of this research.

\section{Conclusion}

We demonstrated a simple method to manipulate mode competition in a modelocked oscillator, based on a controlled combination of the standard homogeneous gain with a small amount of shaped inhomogeneous gain. This combination is very powerfull for precise control over the spectrum of ultrashort pulses within the optical cavity, allowing stable oscillations that are inaccessible with purley homogeneous or inhomogeneous gain. Our concept of intra-cavity gain shaping holds notable advantages over other shaping techniques, either extra- or intra-cavity, as the former are lossy in power and the latter are very limited by effects of mode competition. Intra-cavity gain shaping provides, in a compact single oscillator, flexible, power preserving, so far unattainable control over the emitted pulses, and can generate multi-lobed spectra where the center, width and power of each lobe can be independently set.

\section*{Acknowledgments}
\label{Acknowledgments}

This research was supported by the Israeli science foundation (grant \#807/09) and by the Kahn foundation.


\end{document}